# Cross-Layer Assisted Early Congestion Control for Cloud VR Services in 5G Edge Network

Wanghong Yang[1,2], Wenji Du[1,2], Baosen Zhao[1,2], Yongmao Ren[1,2], Jianan Sun[1], Xu Zhou[1,2]

[1]Computer Network Information Center, Chinese Academy of Sciences, Beijing 100190, China

[2]University of Chinese Academy of Sciences, Beijing 100049, China

{yangwanghong, hjdu, zhaobaosen, renyongmao, jnsun, zhouxu}@cnic.cn

*Abstract*—Cloud virtual reality (VR) has emerged as a promising technology, offering users a highly immersive and easily accessible experience. However, the current 5G radio access network faces challenges in accommodating the bursty traffic generated by multiple cloudVR flows simultaneously, leading to congestion at the 5G base station and increased delays. In this research, we present a comprehensive quantitative analysis that highlights the underlying causes for the poor delay performance of cloudVR flows within the existing 5G protocol stack and network. To address these issues, we propose a novel cross-layer informationassisted congestion control mechanism deployed in the 5G edge network. Experiment results show that our mechanism enhances the number of concurrent flows meeting delay standards by 1.5× to 2.5×, while maintaining a smooth network load. These findings underscore the potential of leveraging 5G edge nodes as a valuable resource to effectively meet the anticipated demands of future services.

*Index Terms*—Cloud VR, 5G Edge Network, Congestion Control, Delay Prediction, Cross-Layer Awareness

## I. Introduction

The cloud virtual reality (VR) technology, consisting of cloud computing and cloud rendering, reduces the requirements for hardware computing power, and can provide users with a more flexible, higher-definition, and richer service experience. To create an immersive interactive experience, cloudVR must support both ultra-wide bandwidth and low interaction delay.

However, the current cloudVR services still have many problems in the 5G edge network. The characteristics of frame-by-frame rendering and unified sending of VR streams determine the extremely high peak-to-average ratio, which means the bursty traffic pattern. Through our analysis, the 5G radio access network (RAN), which is often the bandwidth bottleneck in the end-to-end link, cannot guarantee low transmission delay of concurrent flows, resulting in extremely poor quality of user experience (QoE) and significant fluctuations in channel load.

According to the 5G edge network architecture design, the edge nodes have the ability to quickly obtain the underlying network status information through dedicated signaling tunnels [1]. For cloudVR services with real-time interaction requirements, the edge nodes can collect accurate underlying link information and interact with the client or server efficiently to achieve global and efficient transmission control.

Our contributions can be briefly summarized as follows:

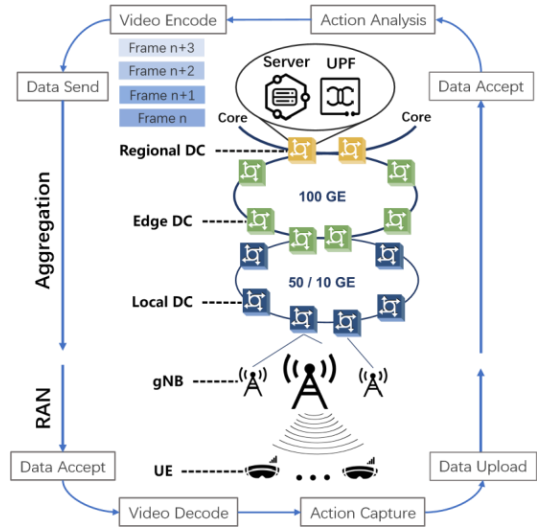

Fig. 1: Transmission process of cloudVR stream on the 5G edge network.

- We analyze the delay problems of concurrent cloudVR flows in existing 5G protocol stacks and architectures by theoretical derivation and simulation evaluation.
- We propose our cross-layer assisted early congestion control mechanism (CECC) which deployed in the 5G user plane function (UPF) unit, providing advanced prediction and proactive schedule to avoid congestion, enabling lowdelay and high-throughput concurrent transmission.
- We confirm the effectiveness and advantages of CECC in network simulator 3 (NS3) comparing with CUBIC and BBR, which represent two types of congestion control algorithms (CCAs) respectively.

We analyze the characteristics of cloudVR in the 5G edge network in Section II. The design of the CECC is described in Section III, followed by the performance evaluation in Section IV. Finally, some conclusions and future work discussion are given in Section V.

## II. Quantitative Analysis and Motivation

Fig. 1 illustrates the transmission process of one cloudVR stream in a 5G edge network formed by an aggregation network and a 5G RAN. The transmission delay that VR users can tolerate from data sending to accepting is minimal. Stripping out processes such as frame rendering and en/decode, the transmission round-trip time (RTT) often needs less than 2030 ms. Therefore, a typical delay standard for cloudVR is that 99.99% of packets should be no more than 10 ms [2], [3], which poses a severe challenge to the 5G edge network.

*A. Delay Performance of CloudVR in 5G RAN*

We start by comparing the instantaneous throughput of cloudVR flows with the capacity of 5G RAN.

The bursty traffic pattern of cloudVR flows: For VR video flows, the average bit rate determines the clarity of the picture, which means the amount of data per second, and the frame rate determines the smoothness of the picture, which means the number of frames per second. Given these two parameters, the size of one video frame can be calculated.

In order to minimize the transmission delay, each rendered frame needs to be transmitted promptly. Therefore, the millisecond-level instantaneous throughput (TH) of the VR flow at the link bottleneck, RAN, can be obtained from the following formula:

$$TH = \frac{SingleFrameSize}{1ms} = \frac{BitRate \times 1s}{FrameRate \times 1ms} \quad (1)$$

According to the QoE requirements of VR users, the frame rate is at least 60 FPS, and the typical bit rate can range from 35 Mbps to 50 Mbps or more. From Equation 1 we know that for a cloudVR flow with an average bitrate of 50 Mbps and a framerate of 60 FPS, the instantaneous throughput reaches 833.3 Mbps.

The limited capacity of 5G RAN: Compared with the 5G core network and aggregation network connected by fibers, the 5G RAN often has smaller and more fluctuating bandwidth, thus becoming the bottleneck of the 5G edge network. The capacity of a 5G base station (gNB) is influenced by multiple factors, notably encompassing the channel bandwidth, the available spectrum, the modulation and coding scheme, and the number of antennas. According to previous measurements by researchers, the average capability of 5G gNB in the FR1 band is around 880Mbps [4].

Given the limited movement range of cloudVR users, the environmental factors typically remain stable, leading to a steady capacity of the gNB. In that case, a single cloudVR flow from the server can cause a heavy load on the 5G RAN. The delay composition of concurrent VR flows: We analyze the processing delay and queue delay with the example of a 2-flow concurrency scenario. Fig. 2(a) depicts the possible traffic pattern of 2 cloudVR flows transmitted simultaneously in the link. Fig. 2(b) depicts the corresponding variation of the downlink one-way delay of one of the flows.

Compared to LTE, 5G RAN utilizes the slot as the basic unit for radio information exchange, offering enhanced flexibility in various scenarios. For instance, in cloudVR scenarios primarily characterized by downlink traffic, the current gNB utilizes a fixed 5 ms frame structure known as DDDSU (where D is downlink subframe, U is uplink subframe, and S is the special subframe comprising a downlink slot, an uplink slot, and a guard period). Once the slot composition of S is determined, the downlink and uplink transmission times of the gNB within a frame can be established.

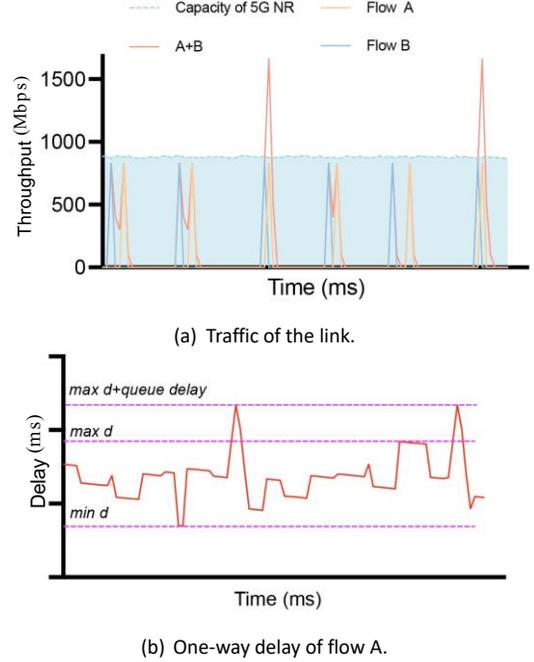

(a) Traffic of the link.

(b) One-way delay of flow A.

Fig. 2: Network traffic and one-way delay in concurrency scenario.

The minimum processing delay is realized when the first packet of a batch of data is sent through the downlink slot immediately. The maximum processing delay is realized when the last packet of a batch of data waits for an uplink time slot before sending.

The queue delay is introduced by the simultaneous arrival of multiple flows. According to the definition of 5G RAN, the gNB with the Single-Input-Single-Output structure establishes separate buffer queues for each flow and poll the data among different flows in radio link control (RLC) layer. Consequently, packets at the tail encounter significant queue delay.

*B. The Choice of Congestion Control*

Since the cloudVR server is deployed in the data center, its specific traffic pattern is also related to the transport layer protocol. Currently, the transport layer protocol used by various cloudVR applications includes UDP, TCP, QUIC, etc. UDP is a best-effort unreliable delivery without a flow control mechanism and used for most video flows today. TCP and QUIC have reliable transmission mechanisms such as congestion control and acknowledgment retransmission.

UDP as transport layer protocol: The cloudVR flows carried on UDP which has no active control mechanism will collide frequently at the gNB, resulting in increased delay and even packet loss.

The traffic pattern shown in Fig. 2(a) is transmitted based on UDP. In best-effort mode, the transport layer protocol does not affect the frame sending rate of VR flows. Since the actual frame spacing of VR flows follows a normal distribution, we can analyze the collision probability of multiple concurrent flows. For a single cloudVR flow, the generation time of the frames follows normal distribution $N(\mu,\sigma^2)$, the traffic is bursty, and the peak rate reaches A Mbps. For N flows to start at the same time, the total peak rate could reach N*A Mbps. We tried to find the time of the next N*A.

Assume $X_{ij}$ represents the sending time of $j_{th}$ frame of flow i. Note that the time for flow i to accumulate to the $T_{th}$ transmission is $Y_{iT} = \sum_{j=1}^{T} X_{ij}$. So $Y_{iT}$ will follow normal distribution $N(\mu*T, \sigma^2*T)$. Then the collision problem can be described as:

$$P(|Y_{maxT} - Y_{minT}| < 1ms) \qquad (2)$$

When $N = 2$, $P(|Y_{maxT} - Y_{minT}| < 1)$ is $P(|X_2 - X_1| < 1)$, which is a two-dimensional normal distribution function. It can be derived that there is a positive probability of next N*A.

When $N>2$, it is no longer normal distribution, but we have confirmed the occurrence of a subsequent collision in our simulation experiments.

TCP/QUIC as transport layer protocol: We first discovered and identified two delay metrics for cloudVR flows, and then divided the CCAs into two categories based on their different performance on these two metrics. Two delay metrics are:

- Frame delay: The time between the frame entering the server transport layer buffer and arriving at the UE transport layer acceptance buffer, which is the total delay experienced by cloudVR users.
- Network delay: The time between the frame being sent from the server node (physical interface) and arriving at the UE network port.

As shown in Fig. 3, the first type of CCAs, represented by BBR, actively detects link delay and bandwidth changes and modifies its transmission rate based on the calculated capacity of the link. BBR smoothes out the instantaneous peak of the millisecond level into a continuous small flow of the second level, avoiding the queuing of a large amount of multi-flow data in the network, thereby achieving a stable and small network delay. However, the tail packet of each frame will face a long wait in the transport layer buffer, resulting in a high frame delay.

The second type of CCAs, represented by CUBIC, includes NewReno, Westwood, and a variety of other algorithms that adjust congestion window (cwnd) by passively sensing changes in packet loss rate or delay. The slow start algorithm needs a certain time to increase cwnd to meet the single frame size. The first two frames need to wait in the buffer, resulting in a high frame delay in the slow start phase. If cwnd is larger than the size of a single frame, CUBIC immediately sends the frame into the sender buffer, thus ensuring that the network delay and frame delay are almost the same.

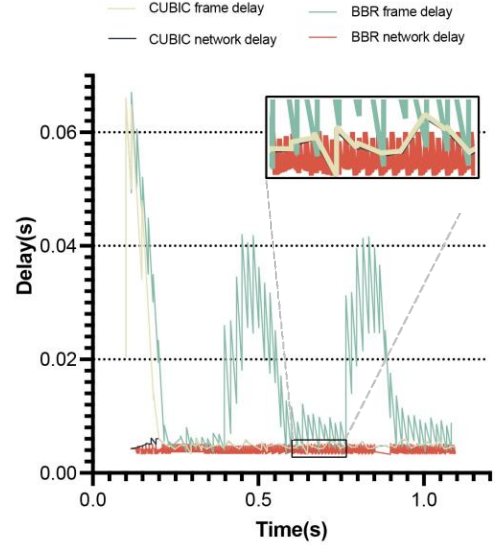

Fig. 3: Comparison of frame delay and network delay of single cloudVR flow carried on different types of CCAs.

*C. Motivation*

A mechanism that can predict the future delay of each flow and schedule them in advance could reduce even avoid congestion. The location of deployment will significantly affect the accuracy of prediction and the foresight of the scheduling. We compared and analyzed state-of-the-art research and combined our needs to determine the best location for early congestion control. Zhuge and Tutti tried to deploy on gNB and edge servers from the perspective of network operators and service providers respectively [5], [6]. However, their work only does moderation for a single flow and lacks a global perspective. The end-side decision algorithm is easier to develop. The disadvantage is that the information is transmitted back with a long delay, reducing the prediction accuracy [7], [8].

UPF would be the most suitable position without changing the end-side protocol. According to the 5G architecture design, a GTP-U tunnel between the UPF and the gNB allows instant information-sharing via separate signaling packets [1]. Meanwhile, the UPF located in the same data center as the server allows global regulation of traffic sent into the network. Therefore, we designed our congestion control mechanism assisted by UPF and gNB.

## III. CECC Design

Our design framework is shown in Fig. 4. With the information provided by the gNB, UPF performs delay prediction for concurrent flows, and schedules the sending of different flows through delay ACK to reduce or avoid network congestion.

### A. Information Report of gNB

The key parameters at the gNB include the queuing information of concurrent flows and the link state information of itself. Queuing information of N flows can be recorded as an array RLC buffer size (*RBS[N]*).

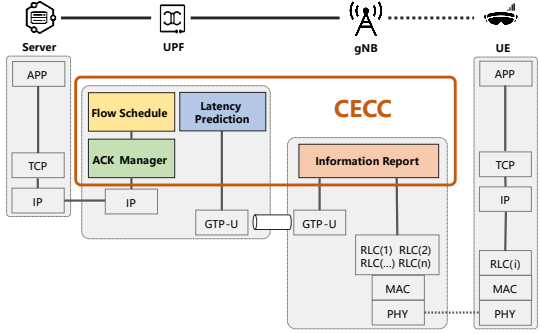

Fig. 4: Framework overview of CECC.

Link state mainly depends on transport block size (*TBSize*) and subcarrier spacing (*SCS*). The former determines the data block's size that is transmitted over the air interface in one slot. The latter determines the number of slots in a subframe and the duration of a slot. With stable bandwidth (*BW*) and modulation coding scheme (*MCS*) in clouVR scenario, the capacity of the gNB can be calculated.

The gNB reports information to the UPF through the GTPU tunnel, where *RBS[N]* is reported in 1 ms intervals and link state information is reported only when there is a change. The reporting adopts the method of carrying multi-flow information in a single-flow single-signaling packet to save packet header overhead.

### B. Delay Prediction

After obtaining the above parameters provided by the gNB, the UPF can use the following formula to predict the future delay of each flow:

$$PreDelay[i] = WiredD + QueueD[i] + ProcessingD \quad (3)$$

Among them, *WiredD* can be obtained through the timestamp of the signaling packet communicated between UPF and gNB. *ProcessingD* is calculated based on the parameters reported by gNB. *QueueD[i]* can be calculated by the following formula:

$$QueueD[i] = \frac{RBS[i] + TxSize[i] + FrameSize[i]}{TH/N} \quad (4)$$

Specifically,

- *RBS[i]* represents the data size already queued in the gNB RLC buffer of flow *i* at the last report.
- *TxSize[i]* represents the data size sent by UPF to gNB since the last report. Here we create separate logs at the UPF interface for each flow to record the timestamp and data size sent at each time.
- *FrameSize[i]* represents the data size to be sent next time for flow *i*. This value is taken to be the maximum value of the data size recorded in the log during a period of smooth fluctuation.
- *TH* represents the current capacity of the gNB.
- *N* represents the number of flows currently queuing in the gNB.

### C. Flow Schedule

As shown in Algorithm 1, UPF records the sending of all flows and reduces the priority of the flows that have just been sent to the lowest. The predicted delay for all flows is then compared to the user delay standard set to 10 ms. Any flow that exceeds the standard means that the current link is overloaded. The UPF kicks the lowest priority flow out of consideration, then re-predicts and re-compares the delay of each flow. The ACKs of the kicked-out flows are cached in the UPF buffer and await computation and schedule for the next cycle.

**Algorithm 1: Flow Schedule**

1. Input: set *N* of flows with ACK in UPF, set *P* of all flows, delay standard
2. Output: set *M* of flows that can be sent in this cycle
3. Initialize priority[*P*] = [0, 0, ..., 0];
4. Function flow monitor(*P*)
5.   //this function keeps running in background
6.   Detect frame passed by UPF of flow $i \in P$;
7.   priority[*P*] ++ ;
8.   priority[*i*] = 0 ;
9.   return priority[*P*];
10. Function flow control(*N*)
11.   //this function runs in cycle
12.   Calculate PreDelay[*N*];
13.   while *maximum(PreDelay[N]) > delay standard* do
14.     choose one flow $j \in N$ with minimal priority;
15.     $N \leftarrow N - j$ // delete *j* from set *N*;
16.     recalculate PreDelay[*N*] ;
17.   $M \leftarrow N$;
18.   return *M*;

## D. ACK Manager

ACK directly controls the pointer position of each flow's send window and the available send window size. After determining the flow set that can be sent in each cycle, UPF returns two ACKs each time. The receiver window (rwnd) of the first ACK is greater than the frame size, ensuring that the server can send out data in time if there is data waiting to be sent. The rwnd of the second ACK is very small to ensure that the server cannot send data by itself before UPF makes the next decision.

Many CCAs support using a single ACK packet to acknowledge multiple previously received packets, resulting in the number of ACK packets being much smaller than the number of packets sent. To ensure UPF has enough ACKs to regulate the sender, the ACK manager of UPF stores the ACK with the largest sequence number received by each flow and uses the currently recorded returned sequence number and the maximum sequence number to generate the intermediate continuous ACK packets. The generated ACK packet only differs from the real ACK packet in the sequence number, so the server will not reject it.

of that flow, and the black dots show the total delay of that flow when queuing occurs under concurrent collisions.

Under the DDDSU frame structure of 5G gNB, the cloudVR single-flow downlink one-way delay fluctuates between 3.25 and 6.1 ms. As the number of flows increases, the underlying processing delay for all flows increases because the average time slot waiting time for data on the base station side is longer for each flow.

However, the queue delay for CUBIC flows rises extremely fast due to the lack of a staggering mechanism. CECC flows predict the congestion caused by multiple flows being sent at the same time and stagger the sending times of each flow by modulating the ACKs, thus achieving a smaller delay increase.

## C. User Satisfaction

Considering the additional delay in the slow start phase, Fig. 7 shows the satisfaction of concurrent flows on CECC and CUBIC with and without counting the slow start phase.

The small initial cwnd (10 segments) of CUBIC results in a single frame not being fully sent in time, increasing the delay. CUBIC takes 3 RTTs to reach the frame size. Combined with the

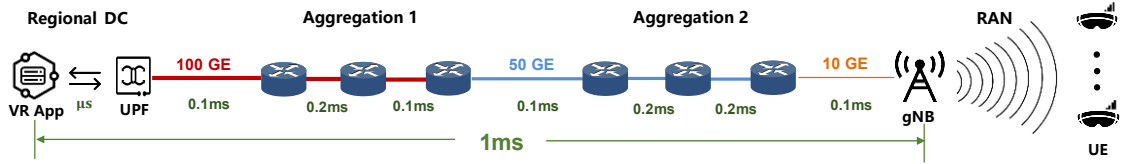

Fig. 5: Experiment topology

## IV. EVALUATION

Based on the analysis in Section II, we compare the delay performance and user satisfaction of concurrent cloudVR flows transmission with and without CECC in NS3, and finally compare the network utilization of concurrent transmission with CECC, CUBIC and BBR.

## A. Experiment Setup

We conducted our experiments by integrating the VR app model into the LENA module of the NS3 [9], [10]. Due to the lack of a MIMO module, we increased the bandwidth to approximate the 880 Mbps capacity of the real gNB. The topology of our experiments is shown in Fig. 5. More specific parameter settings are shown in Table 1.

Table I: 5G RAN Parameters

| Parameters | Values | Parameters | Values |
|---|---|---|---|
| Frequency | 3.5GHz | Bandwidth | 200MHz |
| TxPower | 30dBm | TDD Pattern | DDDSU |
| Subcarrier | 15kHz | Propagation model | UMa LoS |

## B. Delay Comparison

Fig. 6 illustrates the delay reductions of deploying CECC for concurrent cloudVR flows compared to using the original CUBIC. Each box shows the basic processing delay fluctuation processing delay on the UE side, the RTT of each flow fluctuates between 15 and 18 ms. After 3 RTTs, the accumulated volume in the sender's buffer will reach 200 KB, increasing queue delay of about 2 ms in the gNB. As a result, the percentage of packets with one-way delays greater than 10 ms across all flows ranged from 0.11% to 0.14%, unable to meet the standard of less than 0.01% and reducing user satisfaction to 0.

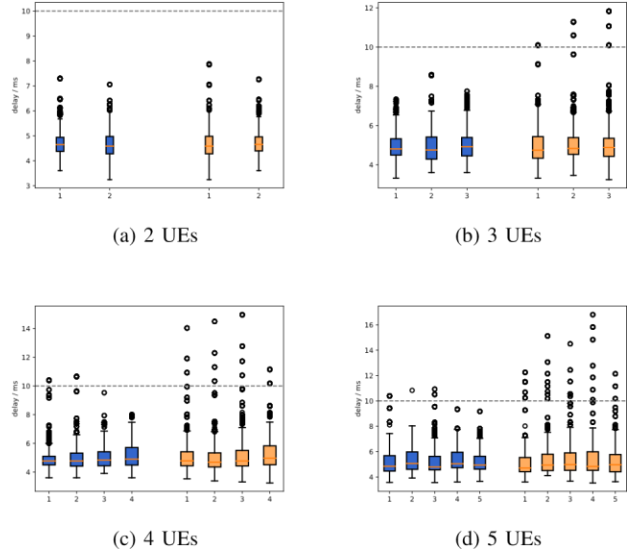

(a) 2 UEs  (b) 3 UEs

(c) 4 UEs  (d) 5 UEs

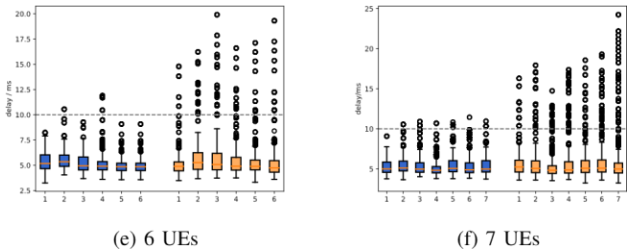

Fig. 6: Delay performance of each flow in CECC (blue boxes) and CUBIC (yellow boxes).

If the effect of slow start is excluded and only the stabilization phase where cwnd is large enough is considered, the delay will increase by about 1 ms on the gNB side due to the average frame size of 100KB for each flow, so the oneway delay for each flow in the stabilization phase at 5-flow concurrency will exceed 10 ms and the number of satisfied users will gradually decrease.

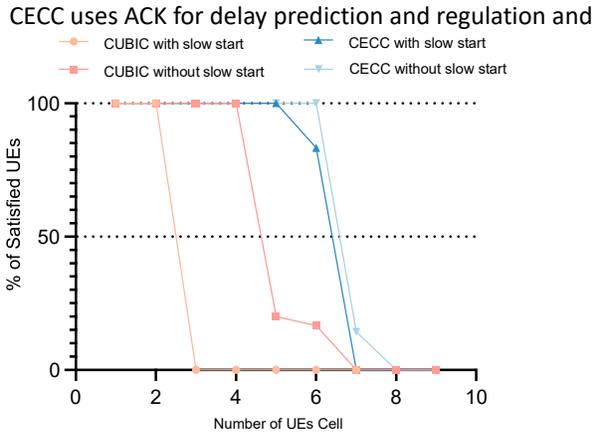

Fig. 7: UEs satisfaction.

can carry out send control for each flow after 1 RTT, so the slow start has less impact on CECC. The maximum percentage of packets with one-way delay greater than 10 ms for 4 and 5 concurrency carried by CECC is 0.006%, which still meets the standard.

For 6 concurrent flows, one flow exceeds the standard due to high delay during the slow start phase, so the satisfaction rate drops from 100% to 83.3%. With 7 flows concurrent, the basic processing delay increased and the increase in queue delay caused by collisions was even greater, resulting in decreased satisfaction as the delay exceeded 10 ms for all flows.

*D. Network Utilization*

Fig. 8 shows the load on the aggregation network with different concurrent flows carried by BBR, CUBIC and CECC. The abscissa axis indicates the total number of bytes transferred every 5 ms at the UPF interface.

BBR makes all flows to be continuously and simultaneously transmitted at a lower rate in the aggregation network. Although the network has the least idle time and the least total load, the delay of the tail packet in each frame is the largest. CUBIC and CECC ensure that each frame can be sent out immediately. As the number of concurrent flows increases, the staggering effect of CECC becomes more pronounced. Compared with CUBIC, CECC can reduce the maximum load of the link, thereby effectively reducing the queue delay of each flow.

## V. Conclusion and Future Work

Our cross-layer assisted early congestion control mechanism actively predicts network status and avoids congestion through information released by gNB, guaranteeing low delay transmission in high concurrency scenarios. Edge network nodes such as UPF and gNB have great potential to participate in enhancing both user QoE and 5G edge network load balance.

For the different growth strategies of different CCAs, we are also looking for a more robust solution to reduce the network delay and frame delay in the slow start period. We will consider reducing the report information and reporting

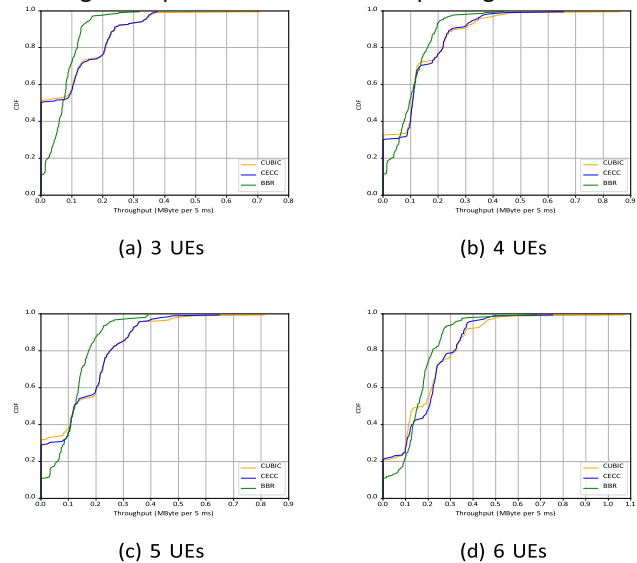

Fig. 8: The CDF of the total throughput of concurrent flows carried by different algorithms.

frequency of the gNB to improve the deployability in our open-source 5G network testbed.

REFERENCES

[1] 3GPP, "3GPP TS 23.288: Architecture enhancements for 5g system (5gs) to support network data analytics services," Technical Specification (TS) 23.288, 3rd Generation Partnership Project (3GPP), December 2019. Version 15.4.0.

[2] Huawei iLab, "Cloud vr solution white paper," white paper, Huawei Technologies Co., Ltd., 2018.

[3] Y. Siriwardhana, P. Porambage, M. Liyanage, and M. Ylianttila, "A survey on mobile augmented reality with 5g mobile edge computing: architectures, applications, and technical aspects," *IEEE Communications Surveys & Tutorials*, vol. 23, no. 2, pp. 1160–1192, 2021.

[4] D. Xu, A. Zhou, X. Zhang, G. Wang, X. Liu, C. An, Y. Shi, L. Liu, and H. Ma, "Understanding operational 5g: A first measurement study on its coverage, performance and energy consumption," in *Proceedings of the Annual conference of the ACM Special Interest Group on Data Communication on the applications, technologies, architectures, and protocols for computer communication*, pp. 479–494, 2020.

[5] Z. Meng, Y. Guo, C. Sun, B. Wang, J. Sherry, H. H. Liu, and M. Xu, "Achieving consistent low latency for wireless real-time communications with the shortest control loop," in *Proceedings of the ACM SIGCOMM 2022 Conference*, pp. 193–206, 2022.

[6] D. Xu, A. Zhou, G. Wang, H. Zhang, X. Li, J. Pei, and H. Ma, "Tutti: coupling 5g ran and mobile edge computing for latency-critical video analytics," in *Proceedings of the 28th Annual International Conference on Mobile Computing And Networking*, pp. 729–742, 2022.

[7] Y. Xie, F. Yi, and K. Jamieson, "Pbe-cc: Congestion control via endpoint-centric, physical-layer bandwidth measurements," in *Proceedings of the Annual conference of the ACM Special Interest Group on Data Communication on the applications, technologies, architectures, and protocols for computer communication*, pp. 451–464, 2020.

[8] M. Chen, R. Li, J. Crowcroft, J. Wu, Z. Zhao, and H. Zhang, "Ran information-assisted tcp congestion control using deep reinforcement learning with reward redistribution," *IEEE Transactions on Communications*, vol. 70, no. 1, pp. 215–230, 2021.

[9] N. Patriciello, S. Lagen, B. Bojovic, and L. Giupponi, "An e2e simulator for 5g nr networks," *Simulation Modelling Practice and Theory*, vol. 96, p. 101933, 2019.

[10] M. Lecci, A. Zanella, and M. Zorzi, "An ns-3 implementation of a bursty traffic framework for virtual reality sources," in *Proceedings of the 2021 Workshop on ns-3*, pp. 73–80, 2021.